%
\input phyzzx
%
%
\overfullrule=0pt
\normaldisplayskip = 15pt plus 7pt minus 7pt
%
%
 
\def\Re{\mathop{\rm Re}\nolimits}

\def\mpl{M_{\rm Pl}}
%
%
\def\ldf{\REF}
\def\JJjournal#1#2{\unskip\space{\sfcode`\.=1000\sl #1 \bf #2}\space }
\def\nup#1 {\JJjournal{Nucl. Phys.}{B#1}}
\def\plt#1 {\JJjournal{Phys. Lett.}{#1}}
\def\cmp#1 {\JJjournal{Comm. Math. Phys.}{#1}}
\def\prp#1 {\JJjournal{Phys. Rep.}{#1}}
\def\prl#1 {\JJjournal{Phys. Rev. Lett.}{#1}}
\def\prev#1 {\JJjournal{Phys. Rev.}{#1}}
\def\mplt#1 {\JJjournal{Mod. Phys. Lett.}{#1}}
%
\ldf\LF{Recent reviews can be found for example in, 
F.\ Quevedo, hep-th/9603074;\brk
J.\ Louis and K.\ F\"orger, 
Nucl.\ Phys.\ B (Proc.\ Suppl.)
{\bf 55B} (1997) 33,  hep-th/9611184.}
\ldf\dienes{For a review see for example,
K.R.\ Dienes, hep-th/9602045.}
\ldf\witten{E.\ Witten, \nup471 (1996), hep-th/9602070.}
%
\ldf\DINDRSW{J.P.~Derendinger, L.E.~Ib\'a\~nez and
  H.P.~Nilles, \plt  155B (1985) 65;\brk
M.~Dine, R.~Rohm, N.~Seiberg and  E.~Witten, \plt  156B (1985) 55.}
\ldf\twogaugino{
N.V.~Krasnikov, \plt  193B (1987) 37;\brk
L.~Dixon, in the proceedings of the A.P.S.
  Meeting, Houston, 1990, ed.\ B.\ Bonner and
  H.\ Miettinen, World Scientific, 1990;\brk
J.~A.~Casas, Z.~Lalak, C.~Mu\~noz and G.G.~Ross, \nup347 (1990) 243;\brk
T.~Taylor, \plt B252 (1990) 59;\brk
B.~de Carlos, J.A.~Casas and C.~Mu\~noz, \nup399 (1993) 623, hep-th/9204012;\brk
R.\ Brustein and P.\ Steinhardt, Phys.\ Lett.\ {\bf B302} (1993) 196, hep-th/9212049.}
\ldf\LEPData{%
Review of Particle Data, Phys.\ Rev.\ {\bf D54}
(1996) 83.}
\ldf\DS{M.~Dine and N.~Seiberg, \plt 162 (1985) 299.}
\ldf\BD{
T.~Banks and M.~Dine, \prev D50 (1994) 7454, hep-th/9406132;\brk
M.~Dine and Y.\ Shirman, Phys.\ Lett.\ {\bf B377}
 (1996) 36, hep-th/9601175;\brk
J.A.\ Casas,
Nucl.\ Phys.\ Proc.\ Suppl.\ {\bf 52A} (1997) 289, 
hep-th/9608010.}
\ldf\vafa{C.~Vafa, Nucl.\ Phys.\ {\bf B469} (1996) 403, hep-th/9602022.}
\ldf\WCYfour{
E.~Witten, Nucl.\ Phys.\ {\bf B474} (1996) 343, hep-th/9604030;\brk
R.\ Donagi,  A.\ Grassi,  E.\ Witten,
Mod.\ Phys.\ Lett.\ {\bf A11} (1996) 2199, hep-th/9607091;\brk
S.\ Kachru and  E.\ Silverstein, Nucl.\ Phys.\  {\bf B482} (1996) 92, hep-th/9608194;\brk
O.J.\ Ganor, hep-th/9612077;\brk
G.\ Curio and D.\ L\"ust, hep-th/9703007.} 
\ldf\CYfour{
M.\ Li, hep-th/9606091;\brk
S.\ Sethi, C.\ Vafa and E.\ Witten, Nucl.\ Phys.\ {\bf B480} (1996) 213, hep-th/9606122;\brk 
I.\ Brunner and  R.\ Schimmrigk, Phys.\ Lett.\ {\bf B387} (1996) 750, hep-th/9606148;\brk 
P.\ Mayr, Nucl.\ Phys.\ {\bf B494} (1997) 489, hep-th/9610162;\brk
I.\ Brunner, M.\ Lynker and R.\ Schimmrigk, Nucl.\ Phys.\ {\bf B498} (1997) 156, hep-th/9610195;\brk 
M.\ Bershadsky, A.\ Johansen, T.\ Pantev, V.\ Sadov and C.\ Vafa, hep-th/9612052;\brk
A.\ Klemm, B.\ Lian, S.-S.\ Roan and S.-T.\ Yau, hep-th/9701023;\brk
M.\ Bershadsky, A.\ Johansen, T.\ Pantev and V.\ Sadov, hep-th/9701165;\brk 
M.\ Kreuzer and H.\ Skarke, hep-th/9701175;\brk
P.\ Candelas, E.\ Perevalov and G.\ Rajesh, hep-th/9704097, hep-th/9707049;\brk
B.\ Andreas, G.\ Curio and D.\ L\"ust, hep-th/9705174.}
\ldf\CS{P.~Candelas and H.~Skarke, hep-th/9706226.}
\ldf\schmal{E.\ Witten, 
Nucl.\ Phys.\ {\bf B460} (1996) 541, hep-th/9511030.}
\ldf\HW{P.\ Horava and E.\ Witten, 
Nucl.\ Phys.\ {\bf B460} (1996) 506, hep-th/9510209; \brk
Nucl.\ Phys.\ {\bf B475} (1996) 94, hep-th/9603142.}  
\ldf\Mgaugino{
T.~Banks and M.~Dine, Nucl.\ Phys.\ {\bf B479} (1996) 173, hep-th/9605136;\brk
E.\ Caceres, V.S.\ Kaplunovsky and I.M.\ Mandelberg, Nucl.\ Phys.\ {\bf B493} (1997) 73,  hep-th/9606036;\brk
P.\ Horava, Phys.\ Rev.\ {\bf D54} (1996) 7561, hep-th/9608019;\brk
A.\ Lukas,  B.A.\ Ovrut and  D.\ Waldram, hep-th/9611204\brk
P.\ Binetruy,  M.K.\ Gaillard,  Y.-Y.\ Wu, hep-th/9702105;\brk
H.P.\ Nilles and S.\ Stieberger, hep-th/9702110;\brk
E.~Dudas and C.~Grojean, hep-th/9704177;\brk
I.\ Antoniadis and M.\ Quiros, hep-th/9705037;\brk
H.P.\ Nilles, M.\ Olechowski and M.\ Yamaguchi, hep-th/9707143;\brk
Z.\ Lalak and  S.\ Thomas, hep-th/9707223
.}
\ldf\DMW{M.J.\ Duff, R.\ Minasian and 
E.\ Witten,
Nucl.\ Phys.\ {\bf B465} (1996) 413, hep-th/9601036.}
\ldf\AM{P.\ Aspinwall and D.R.\ Morrison, hep-th/9705104.}
\ldf\ADS{I.~Affleck, M.~Dine and N.~Seiberg,  \nup241 (1984) 493, \nup256 (1985) 557.}
\ldf\DN{M.\ Dine and A.E.\ Nelson, Phys.\ Rev.\ {\bf D48} (1993) 1277, hep-ph/9303230.}
\ldf\Tbreak{
A.~Font, L.E.~Ib\'a\~nez, D.~L\"ust and F.~Quevedo,
\plt  B245 (1990) 401;\brk
S.~Ferrara, N.~Magnoli, T.~Taylor and G.~Veneziano,
\plt B245 (1990) 409;\brk
H.~P.~Nilles and M.~Olechowski, \plt B248 (1990) 268;\brk
P.~Bin\'etruy and M. K. Gaillard, \nup358 (1991) 121.}
\ldf\IL{
L.E.\ Ibanez and  D.\ L\"ust, Nucl.\ Phys.\ {\bf B382} (1992) 305, hep-th/9202046.}



\Pubnum{hep-th/9708049\cr UTTG--22--97}
\date{}
\titlepage
\title{Phenomenological Aspects of 
         F-theory\foot{Research supported in part by:
	the NSF, under grant PHY--95--11632 (V.~K.);
	the Robert A.~Welch Foundation (V.~K.);
        the German--Israeli Foundation 
        for Scientific Research (J.~L.);
	the NATO, under grant CRG~931380 (both authors).}}
\author{Vadim Kaplunovsky
	\foot{Email: \tt vadim@bolvan.ph.utexas.edu}}
\address{Theory Group, Dept.~of Physics, University of Texas\break
	Austin, TX 78712, USA}
\andauthor{Jan Louis
	\foot{Email: \tt jlouis@hermes.physik.uni-halle.de}}
\address{Martin-Luther-Universit\"at Halle-Wittenberg,\break
Fachbereich Physik, D-06099 Halle, Germany}
\vfil\vfil
\abstract
Stabilizing a heterotic string vacuum with a large expectation value
of the dilaton and simultaneously breaking low-energy
supersymmetry is a long-standing problem of string phenomenology.
We reconsider these issues in light of the recent developments in
F--theory.
\endpage
{\catcode`\@=11 \global\lastf@@t=-1 }


Since the inception of the heterotic string theory,
a very large number of phenomenological string models were constructed
based upon various candidate ground states of the string.
Although such models differ from each other in countless details,
certain ground rules have been firmly established as either universal
consequences of the perturbative heterotic string theory or else
essential for obtaining the correct Standard Model phenomenology.\refmark{\LF}
Generally, a string model has four spacetime dimensions,
${\cal N}=1$ supersymmetry and a large gauge symmetry $G=\prod_a G_a$
including the $SU(3)\times SU(2)\times U(1)$
of the Standard Model as well as additional, `hidden' factors.
At the tree level, all the gauge couplings $g_a$ are controlled by
the expectation value of the dilaton field $S$,
$$
{4\pi\over g_a^2}\,\equiv\,{1\over\alpha_a}\
=\ k_a\,\vev{\Re S} 
\eqn\gtree
$$
($k_a$ being fixed integer or rational coefficients).
The universality of this relation
naturally leads to the desired  GUT-like pattern of the Standard Model's
gauge couplings.
\foot{%
    The string analogue of the GUT  scale $3\cdot 10^{17}$~GeV
    does not exactly coincide with the phenomenologically favorite value
    $M_{\rm GUT}\approx 2\cdot 10^{16}$~GeV, but the discrepancy is small
    enough to be explainable (in principle) in terms of the perturbative
    string threshold corrections.\refmark{\dienes}
    It has also been suggested that this mismatch
    can disappear in strongly coupled heterotic vacua.\refmark{\witten}
    }
The perturbative string theory suffers from an exact degeneracy which leaves
$\vev S$ completely undetermined;
likewise, the vacuum expectation values
of several other moduli fields (collectively denoted $T$) are
also indeterminate to all orders of the string perturbation theory.

The lifting of this degeneracy 
as well as spontaneous supersymmetry breakdown
can be accomplished with the help of field-theoretical
non-perturbative effects arising from asymptotically free (and hence
infrared-strong) hidden sectors of the gauge group.\refmark{\DINDRSW}
In the simplest scenario, a confining hidden sector 
generates a dynamical
superpotential $W\sim\Lambda^3_{\rm hid}$ where
$$
\Lambda_{\rm hid}\ \sim\ e^{-2\pi k S/b}\,\mpl
\eqn\LambdaHid
$$
is the confinement scale
 and $b$ the appropriate $\beta$-function coefficient. 
Taking several such hidden sectors together and allowing for moduli-dependent
pre-exponential factors, one generally has\refmark{\twogaugino}
$$
W_{\rm eff}(S,T)\ =\ \mpl^3\sum_a
C_a(T)\,e^{-6\pi k_a S/b_a}
\eqn\Weffective
$$
($a$ runs over the confining hidden sectors),
which leads to an effective scalar potential 
$V(S,T)=e^K\bigl[\left|DW\right|^2-3\left|W\right|^2\bigr]$.
Phenomenologically, this effective potential should have  
a stable minimum with spontaneously broken supersymmetry
and zero cosmological constant.
Furthermore, the observable sector (\ie, the Standard Model)
should feel the breakdown of supersymmetry at the 
electroweak scale $M_W$;
this requires $W_{\rm eff}=O(M_W\mpl^2)$
or equivalently  confinement scales $\Lambda_{\rm hid}$
in the $10^{13}$~GeV to $10^{14}$~GeV range.

In all other scenarios, the hierarchy
$M_W\ll\mpl$ also follows from $\Lambda_{\rm hid}\ll\mpl$
for some kind of a hidden sector.
According to eq.~\LambdaHid, this requires a rather large expectation
value of the dilaton field, typically $\vev{Re S}\gsim 10$ or more.
Likewise, extrapolating the Supersymmetric Standard Model
all the way up to the GUT scale and using eq.~\gtree, one needs 
$\vev{Re S}\approx\alpha_{\rm GUT}^{-1}\approx23$.\refmark{\LEPData}
Unfortunately, it is extremely difficult to stabilize the dilaton at
such a large value using only field-theoretical non-perturbative effects.
According to Dine and Seiberg,\refmark{\DS}
for any string model with
unbroken supersymmetry at the tree level, 
the effective potential exponentially
asymptotes to zero in the weak coupling regime $\Re S\to\infty$ and
hence, the stable minima of the potential, if any, must lie at strong coupling.
\foot{See also the discussion in refs.~[\BD].}
For example,  the superpotential \Weffective\ with
a generic K\"ahler function $K(S,T)$ and no special tuning of the
coefficients $C_a(T)$ and $k_a/b_a$,
leads to stable vacua only when some
of the exponential factors $e^{-6\pi k_a S/b_a}$ are not too small ($O(1)$)
and hence
$$
\vev{\Re S}\ \lsim\ O(1/6\pi)\,\max_a (b_a/k_a)\,.
\eqn\dilalimit
$$
An example of a hidden sector with a large $b/k$ ratio is the
unbroken second $E_8$ gauge group of the heterotic string ($b=90$, $k=1$);
depending on how seriously one takes the numerical factors in eq.~\dilalimit,
it might be barely consistent with
stable $\vev{Re S}\approx23$.
At the same time however, one would have $\Lambda_{E_8}=O(\mpl)$,
which leads
to spontaneous supersymmetry breakdown close to the Planck scale
and hence no hierarchy.

From the purely field-theoretical point of view, one might 
attempt to solve the problem by employing several large
hidden gauge factors\refmark{\twogaugino} to stabilize
a large $\vev{\Re S}$ without breaking supersymmetry
and then add yet another hidden sector with $\Lambda\ll\mpl$ for the express
purpose of breaking supersymmetry at a hierarchically small scale.
\foot{%
    The Krasnikov mechanism --- stabilizing a large $\vev{\Re S}$ by
    using two or more hidden gauge factors ---
    can be fine-tuned to work with modestly
    sized hidden gauge group provided their $\beta$-functions are very
    close but not exactly equal, $b_1\approx b_2\gg|b_1-b_2|>0$.
    With the help of the moduli fields $T$ --- and even more fine tuning
    --- one may achieve a hierarchical supersymmetry breakdown.
    However, the extreme fine tuning required by this scenario makes
    it rather marginal for model-building purposes.
    }
However, from the heterotic string's point of view,
this scenario ---
or any other scenario which needs very large or complicated hidden sectors
--- conflicts with the universal central charge constraint,
which limits the rank of the entire (perturbative)
four-dimensional gauge group:
$$
{\rm rank}(G) \le 22\,;
\eqn\bcc
$$
this leaves rather limited room for the hidden sectors.
Consequently, the perturbative heterotic string theory
with only field-theoretical non-perturbative corrections
has extreme difficulty combining a stable vacuum with
a large dilaton expectation value and a large hierarchy.

The inherently stringy non-perturbative effects are now gradually
becoming understood in terms of duality relations between various
string theories, M--theory and F--theory.
In particular, the ${\cal N}=1$, $d=4$ compactifications of the
heterotic string are dual to F--theory compactifications on
elliptically fibered Calabi-Yau fourfolds.\refmark{\vafa,\WCYfour,\CYfour,\CS}
This duality shows that the perturbative heterotic string theory 
often reveals only a small part of the ultimate four-dimensional gauge
group $G$ while many additional gauge fields arise from
singularities 
of the heterotic compactification where the perturbation theory
breaks down.\refmark\schmal~\strut
\foot{%
    This is distinct from the perturbation theory breakdown due
    to a large overall ten-dimensional
    heterotic string coupling.
    That regime is best described in terms of the dual M--theory
    \refmark{\HW};
    some of its phenomenological implications are discussed
    in refs.~[\witten,\Mgaugino].
    }
Only the perturbative gauge couplings $g_a$ are governed by the dilaton $S$;
the non-perturbative couplings are dilaton-independent
and are instead controlled by some combinations of the moduli fields $T$.\refmark{\DMW}
From the F--theory point of view, however, all the gauge groups $G_a$
have equal status and $S$ is no different from the other moduli fields.
Only in the corner of the F-theory moduli space which is dual to
the weakly coupled heterotic string does the $S$ field acquire its special
properties.
There are no known constraints on the size or the variety of the
non-perturbative gauge symmetries.\refmark{\AM}
The current record holder among the F--theory compactifications
has 251 simple gauge group factors of total non-abelian 
rank of 302896, the biggest factors being $SO(7232)$ and $Sp(3528)$.\refmark{\CS}

What then are the implications of this non-perturbative
bounty of gauge fields for the string model building?
On one hand, we no longer have any general constraints --- or guidelines
--- for the hidden sectors of string models.
On the other hand, it is precisely the absence of constraints such as \bcc\
that makes it easy to obtain a small $\alpha_{\rm GUT}$ in a stable vacuum.
For example, imagine a model where 
the same combination $\tilde S$ of moduli fields
governs the gauge couplings of the Standard Model and also of several
{\it large} confining hidden gauge groups.
In this model, the effective superpotential for  $\tilde S$ and the other
moduli looks exactly like \Weffective\ (modulo replacement $S\to\tilde S$)
and hence
the minima of the resulting effective potential are generally
found at $6\pi \vev{\Re\smash{\tilde S}}= O(b_{\rm hid})$.
This is compatible with $\Re\tilde S\approx 23$ for
$b_{\rm hid}\sim 400$ --- as in \eg, pure-gauge $SO(140)$ ---
which would be easily obtainable in F--theory compactifications
(but quite out of reach of the perturbative heterotic theory). 

In a more realistic model, different gauge couplings $\alpha_a(T)$
may be controlled
by different dilaton-like combinations of the moduli fields $T$
(which by abuse of notations now include the heterotic $S$ field as well).
In this case, one has
$$
W_{\rm eff}(T)\ =\ \mpl^3
\sum_a C_a(T)\,e^{-6\pi/b_a\alpha_a(T)}
\eqn\Wgeneral
$$
or even a more complicated non-linear combination of the
$e^{-6\pi/b_a\alpha_a(T)}$
if there are hidden matter fields that are charged under several hidden
gauge groups at once.
Generally, there are also inherently F--theoretical instantonic contributions,
although they are believed to be smaller than those of confining
hidden sectors.\refmark{\WCYfour}
The precise behavior of the resulting scalar potential
can only be analyzed on the model-by-model basis, but a crude
order-of-magnitude analysis suggests that
its stable minima (if any) should have
$$
{6\pi\over\alpha_{\rm hid}(\vev{T})}\ =\ O(b_{\rm hid}) .
\eqn\Ageneral
$$
Again, we see that large hidden sectors naturally lead to small
$\alpha_{\rm hid}\)$.

In a generic F--theory compactification, however, small
$\alpha_{\rm hid}$ do not necessarily imply a weakly coupled
Standard Model.
Furthermore, there is no longer an automatic GUT-like unification
of the Standard Model's couplings themselves.
Both of these features --- which perturbatively followed from eq.~\gtree\ ---
now have to be imposed as phenomenological
constraints on the F-theory models.
Specifically, the three Standard Model's couplings
should be governed by the same modulus,
\foot{%
    This suggest that in F--theory one should seriously
    consider a possibility of an actual field-theoretical 
    Grand Unification of the $SU(3)\times SU(2)\times U(1)$
    into a simple gauge group with a single $\alpha_{\rm GUT}(T)$.
    }
which is also involved with the hidden gauge couplings~\Ageneral\
and thus obtains a large expectation value.

Notice that eq.~\Ageneral\ implies $\Lambda_{\rm hid}\sim\mpl$
and hence $W_{\rm eff}\sim\mpl^3$.
Therefore, it is imperative that the resulting effective potential
does {\sl not} lead to spontaneous supersymmetry breakdown ---
otherwise, supersymmetry would be broken right at the Planck scale
for all sectors of the theory and there would be no hierarchy.
Likewise, we do not want a Planck-scale cosmological constant.
This gives us two more phenomenological requirements that any
viable F--theory model must satisfy.
Mathematically, these requirements amount to a constraint on 
the holomorphic $W_{\rm eff}$,
\foot{%
    Specifically, one needs a simultaneous solution of holomorphic
    eqs.~$W_{\rm eff}=\partial_T W_{\rm eff}=0$, which
    is automatically a local minimum of the scalar
    potential with unbroken supersymmetry and zero cosmological
    constant.
    Whether or not it is the global minimum of the potential
    depends on the K\"ahler function.
    }
which (in principle) allows one to decide the viability of
any particular model
in terms of exactly computable quantities.

Having survived the Planck-scale physics unbroken, supersymmetry
should be eventually broken down at a hierarchically lower scale.
This can be accomplished by additional hidden sectors
--- fortunately, they are easily available in F--theory.
Such a sector needs the following features:
A week coupling $\alpha\sim\alpha_{\rm GUT}$,
a modest amount of asymptotic freedom $b\sim10$ ---
which together provide for
the hierarchy $\Lambda\ll\mpl$ ---
and most importantly, supersymmetry-breaking infrared dynamics.
\refmark{\ADS}
There are several possibilities for such
supersymmetry breakdown and for the way it affects
the Standard Model.
In the simplest scenario, vacuum stabilization and supersymmetry
breaking result from completely separate hidden sectors:
Large gauge groups (\eg, pure-gauge $SO(140)$) have $\Lambda\sim\mpl$
and create $W_{\rm eff}$ that stabilizes {\it all} the moduli $T$
of the F--theory right at the Planck scale,
while a  sector such as $SU(5)$ with $\bf 10+\bar 5$ matter has
$\Lambda$ in a multi-TeV range and breaks supersymmetry dynamically
without any help from the moduli fields or supergravity.\refmark{\ADS}
Finally, an abelian hidden gauge field communicates the supersymmetry
breaking to the Standard Model {\it \`a la} Dine-Nelson.\refmark{\DN}
Alternatively, the feed-down of supersymmetry breaking to the Standard
Model can proceed through the supergravity- or moduli-mediated
interactions.
Also some of the moduli fields can avoid getting Planck-scale masses
and survive to participate in the supersymmetry-breaking process.
\foot{%
    In this scenario, $\Lambda_{\rm hid}$ 
    has to be $10^{13}-10^{14} GeV$ in order to generate
    an effective superpotential
    for the surviving moduli fields, and it is this superpotential
    that leads to the spontaneous supersymmetry breakdown.
    This is rather similar to the heterotic toy models in which a
    large $\vev S$ is fixed by hand, while the spontaneous supersymmetry
    breakdown is induced by the effective potential for the $T$ modulus.
    \refmark{\Tbreak}
    }
Unfortunately, 
any involvement of the moduli fields in supersymmetry breaking
is likely to destroy
the charge universality of the squark and slepton masses
of the Supersymmetric Standard Model.\refmark{\IL}
From this point of view, the Dine--Nelson scenario appears more attractive.

Let us now summarize the key points:
The non-perturbative string theory or F--theory allow for essentially
unilimited hidden sectors.
This makes it relatively easy to arrange for a stable vacuum state
where supersymmetry is broken at a hierarchically low scale.
On the other hand, the GUT-like unification of the Standard Model's
gauge couplings is no longer automatic but instead has to be imposed
as a phenomenological constraint.
We do not propose any specific models but
merely outline a general scenario for obtaining viable phenomenology
from the F--theory.
Indeed, it is hard to be specific without a better understanding of the
moduli dependence of the gauge couplings in F--theory or even general
rules for obtaining the spectra of the charged matter fields.
However, we believe our scenario is a useful starting point for future work.

\par\smallskip
\noindent {\bf Acknowledgements: }
We would like to thank the CERN theory
group for hospitality and P.\ Mayr for helpful discussions.
\subpar
The research of V.~K.\ is supported in part by the NSF,
under grant PHY--95--11632,
and by the Robert A.~Welch Foundation.
J.~L.\ is supported in part by the German--Israeli 
Foundation  for Scientific Research.
The collaboration of the two authors is 
additionally supported by the NATO, under grant CRG~931380.

\refout
\bye